\documentclass[twocolumn,showpacs,amsmath,amssymb,pra,longbibliography%,superscriptaddress
]{revtex4-1}

\usepackage[pdftex]{graphicx}
\usepackage[usenames,dvipsnames]{color}
\usepackage{epstopdf}
\usepackage{hyperref}
\usepackage{tabularx}

\newcommand{\p}[1]{\mathbf{p}}
\newcommand{\q}[1]{\mathbf{q}}%momenta

\newcommand{\PP}[1]{\mathrm{(L)}}%particle/anti-particle
\newcommand{\AP}[1]{\mathrm{(R)}}

\newcommand{\w}[1]{\mathbf{\hat w}}

\newcommand{\z}[1]{\theta}
\newcommand{\mix}[1]{\Omega}

\newcommand{\aop}[1]{\hat{a}}
\newcommand{\cop}[1]{\hat{c}}

\newcommand{\sm}[1]{\mathrm{SM}}
\usepackage{slashed}

\graphicspath{{../FIGs/}}

\begin{document}

\title{
Reproducing sterile neutrinos and the behavior of flavor oscillations\\
with superconducting-magnetic proximity effects
}
\author{Thomas E. Baker}
\affiliation{Department of Physics \& Astronomy, California State University, Long Beach, CA 90840}
\affiliation{Department of Physics \& Astronomy, University of California, Irvine, CA 92697}
\date{\today}


\begin{abstract}
The physics of a superconductor subjected to a magnetic field is known to be equivalent to neutrino oscillations.  Examining the properties of singlet-triplet oscillations in the magnetic field, a sterile neutrino--shown to be a Majorana fermion--is suggested to be represented by singlet Cooper pairs and moderates flavor oscillations between three flavor neutrinos (triplet Cooper pairs).  A superconductor-exchange spring system's rotating magnetization profile is used to simulate the mass-flavor oscillations in the neutrino case and the physics of neutrino oscillations are discussed.  Symmetry protected triplet components are presented as weak process states. Phases acquired due to the Fulde-Ferrell-Larkin-Ovchinnikov effect produce a complex phase that may be responsible for charge-parity violation in flavor oscillations.
\end{abstract}

\maketitle
%\tableofcontents

%%%%%%%%%%%%%%%%%%%%%%%%%%%%%%%%%%%%

\section{Introduction}

Neutrinos \cite{pauli1978letter,cowan1956detection} exist in three known flavors \cite{danby1962observation,kodama2001observation} corresponding to each generation in the Standard Model  \cite{sheldonNP61,weinbergPRL67,salam1968elementary} and have very small masses with respect to other particles \cite{fukudaPRL98,ahmadPRL01,ahmadPRL02}.  These flavor states may oscillate \cite{pontecorvoJETP68,harmerPRL68} between each other.  Measuring the chirality of the neutrino shows that they are always left-handed \cite{wuPR57,leePR56}.  Despite these experimental facts, a complete physical picture is difficult to determine due to the small cross section for measuring neutrino events.  The conventionally accepted Standard Model does not account for oscillations of neutrinos \cite{peskin1995introduction}. 

In certain situations, a condensed matter system may possess properties that allow it to mimic a particle physics system.  For example, dispersion relations causing electrons to obey the Dirac equation \cite{zhangNP05}, properties of Weyl fermions in semi-metals \cite{herringPR37,murakamiNJP07,xuS15,huangNC15,wengPRX15}, topological Majorana modes in gapped proximity systems \cite{nadjS14}, and Anderson-Higgs modes in superconductors \cite{nambuPR60,goldstoneINC61,andersonPRL63,sooryakumarPRL80,littlewoodPRL81,littlewoodPRB82} all mimic the physics found in large particle experiments at high energies \cite{higgsPRL64,englertPRL64,guralnikPRL64,higgsexp,weylZP29}.  Studying the condensed matter physics may allow for details unavailable in the particle physics case to be analyzed in greater detail and with direct experimental verification.

\begin{figure}[rb]
\includegraphics[width=.8\columnwidth]{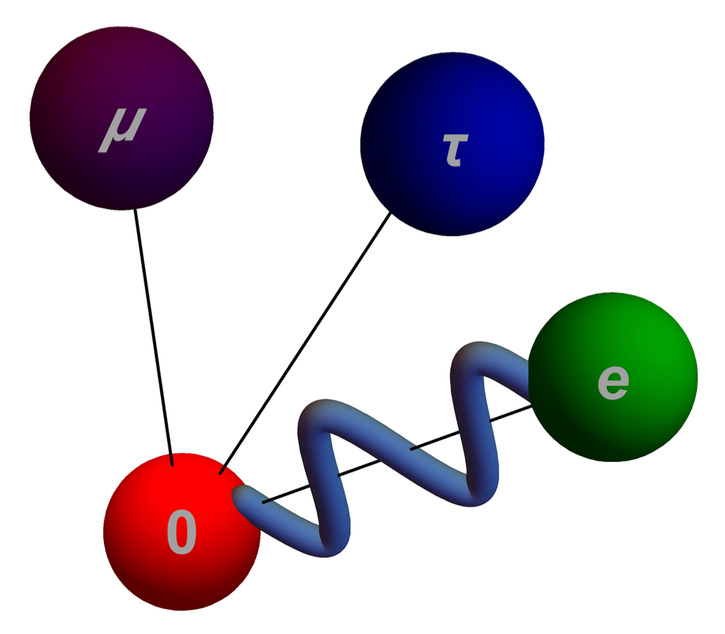}
\caption{A sterile neutrino (0) can transition to a flavored state ($\nu_\mu,\nu_\tau,\nu_e$) just as singlets transition to triplets in superconducting proximity effects.
\label{marionette}}
\end{figure}

A model was constructed by Pehlivan-Balantehkin-Kajino-Yoshida (PBKY) in Ref.~\onlinecite{pehlivanPRD11} which shows that a neutrino gas is analogous to a magnetic field applied to a superconductor.  Interestingly, conventional superconductivity, characterized by the Bardeen-Cooper-Schrieffer (BCS) \cite{BCS} model, is a competing order to magnetism because an electron's spin tends to align with a magnetic field.  This breaks Cooper pairing--which are electrons of opposite spin forming a quasi-particle responsible for the superconducting state \cite{cooper1956bound,BCS,eliashberg1960interactions}. The PBKY model establishes that neutrino oscillations are equivalent to Cooper pairs in the analogy.

Investigating the coexistence of these competing phases of matter is an worthwhile topic of its own; applying a magnetic field can have a variety of effects on the superconducting state depending on its strength.  Weak magnetic fields are expelled from a superconductor \cite{meissnerN33} since the photons of the applied field acquire a finite mass in the symmetry broken superconducting state \cite{peskin1995introduction}.  At high magnetic fields, the superconducting state is destroyed completely or forms an Abrikosov lattice of flux vortices in a type-II superconductor \cite{abrikosov1957magnetic}. 

An intermediate regime of moderate magnetic fields exists where the magnetic field is not strong enough to break the paired electrons and hybridizes the up and down electron bands.  Effectively, the electron with a spin collinear to the magnetic field has an increased momentum (and decreased momentum for the anti-parallel spin).  This momentum splitting gives an oscillation in the superconducting order parameter between the singlet and triplet Cooper pairs and was demonstrated nearly simultaneously by Fulde-Ferrell \cite{fuldePRL64} and Larkin-Ovchinnikov \cite{larkinJETP65} (FFLO).

One system to study the FFLO effect, and therefore neutrino oscillations, is a superconductor in proximity to a ferromagnet \cite{golubovRMP04,buzdinRMP05,bergeretRMP05,chikazumi1978physics,bozorth03}. The superfluid can tunnel into the adjacent material, and cause the entire system to become superconducting \cite{deutscher1969superconductivity} with measurable results at nanoscale distances \cite{ryazanovPRL01,kontosPRL02,keizerN06,khairePRL10}.

The motivating feature of the FFLO effect that begs comparison with neutrinos is that three triplet Cooper pairs are connected to the singlet Cooper pair, just as three flavored neutrinos are connected to a sterile neutrino--a simple extension of the Standard Model to include flavor oscillations \cite{dodelsonPRL94,drewesIJMPE13}--and is summarized in Fig.~\ref{marionette}. A sterile neutrino may couple to the flavor states via the mass term as though it were a right handed particle.  These particles have been suggested to be a candidate for dark matter \cite{zwickyAPJ37,dodelsonPRL94,abazajianPRD01,abazajian2012light}.  The particular sterile neutrino appearing here strongly satisfies the effects required of a dark matter candidate \cite{canettiPRL13}.

In this paper, the connection between BCS superconductivity with an applied magnetic field and the PBKY model from Ref.~\onlinecite{pehlivanPRD11} is reviewed in Sec.~\ref{PBKY}. It is shown that a real external field is applied to the superconducting state in Sec.~\ref{meanfield} by examining the model for flavor states. 

Section~\ref{proximity} provides an overview of superconducting proximity effects by first discussing transport equations of the superfluid.  The physics of those equations is discussed in Sec.~\ref{oscillations}.  

The connection between quantities in the particle physics case and the proximity system are covered in Sec.~\ref{correspondence}.  Section~\ref{ludersexp} identifies neutrino type based on the expansion of the Gor'kov function in the condensed matter case.  Section~\ref{majorana} uses the symmetry of the singlet state to identify the sterile neutrino as a Majorana fermion. The minimally extended Standard Model (MESM) is shown to be analogous to Ginzburg-Landau (GL) theory \cite{landauZETP50} in Sec.~\ref{MESM}.  The physics of flavor oscillations are investigated by comparison with superconductor-exchange spring systems (discussed in Sec.~\ref{SXSS} and further discussed with particular regard to sterile neutrinos in Sec.~\ref{f0role}). Symmetry protected triplet components introduced in Refs.~\onlinecite{bergeretPRL01} and \onlinecite{kadigrobovEPL01} are interpreted as weak process states allowing for the conservation of energy and momentum in Sec.~\ref{weakstates}. The angular momentum quantum number in the superconductor is discussed in the neutrino case in Sec.~\ref{leptonnum}. The Mikheyev-Smirnoff-Wolfenstein (MSW) effect is related to polarization effects in Sec.~\ref{MSW} by noticing similar behavior in the condensed matter case. A possible mechanism for charge-parity (CP) violation from the FFLO phase is discussed in Sec.~\ref{CPviolation}. Other possibilities are covered in Sec.~\ref{exotics}.

\begin{table*}
\begin{minipage}{2\columnwidth}
\begin{center}
\caption{A correspondence table between neutrinos and Anderson's reformulation of BCS superconductivity \cite{andersonPR58} derived from the mathematical analogy of the PBKY model in Ref.~\onlinecite{pehlivanPRD11}.  Items below the divide are introduced in this paper based on the PBKY model.  Note that $\epsilon_\p{}$, the energy of the Cooper pair, is $2\omega$ in the neutrino case. The vector Gor'kov function possesses an extra angular momentum, $\ell$, whose physical significance for the neutrino case is discussed in Sec.~\ref{leptonnum}.  All other connections are justified in Sec.~\ref{correspondence} as well. \label{analog}}
\newcolumntype{C}{>{\centering\arraybackslash}X}
\begin{tabularx}{\textwidth}{| C | c || c | C |}
\hline
 Condensed Matter & Superconductor & Neutrino & Particle Physics\\
 \hline
\hline
Pairing potential&$\Delta(\mathbf{r})$&$G_F/V$&Vacuum symmetry breaking per gas volume\\
Cooper pair momentum&$\mathbf{p}$&$\mathbf{p}$&Momentum\\
Time reversed counterpart&$\epsilon_{\p{}}<0$ or $\omega<0$&$|{\bar\nu}_{e,\mu,\tau}\rangle$&Anti-neutrinos\\
Matsubara frequency ($T\neq0$)&$\omega_n$&$\omega$&Vacuum oscillation\\
Temperature&$T$&$T$&Temperature\\
Gor'kov Function & $\mathcal{F}=f_0+\mathbf{\hat v}\cdot\mathbf{f}$ & $|\nu_\alpha\leftrightarrow\nu_\beta\rangle$ & Neutrino oscillations \\
\hline
\hline
Vector Gor'kov function&$\mathbf{f}$&$|\nu_{e,\mu,\tau}\rangle$&Flavored neutrino oscillations\\
Singlet Gor'kov function&$f_0$&$|\nu_0\rangle$&Sterile Neutrino (Majorana)\\
Magnetic field&$\mathbf{B}$&$B$&Source field\\
Perpendicular, parallel component \cite{bakerPRB16} & $f_\perp,f_\parallel$ & $|\nu_1\rangle,|\nu_\alpha\rangle$ & mass, flavor eigenstate\\
% component & $f_\parallel$ & $|\nu_\alpha\rangle$ & flavor eigenstate\\
Acquired angular momentum&$\ell$&$\ell$&Flavor-Angular Momentum\\
Cartesian coordinates&$\mathbf{\hat r}$&$e,\tau,\mu$&Lepton flavors\\
\hline
\end{tabularx}
\end{center}
\end{minipage}
\end{table*}

\section{Neutrino oscillations}\label{neutosc}

The PBKY model presented in Ref.~\onlinecite{pehlivanPRD11} equates a neutrino gas with a superconductor in a magnetic field.  This is summarized in Sec.~\ref{PBKY}.  The PBKY model is phrased in Sec.~\ref{meanfield} as a mean field of another field and it is shown that the flavor basis has a source field.

\subsection{Overview of the Pehlivan-Balantekin-Kajino-Yoshido model for neutrinos}\label{PBKY}

The most general Hamiltonian we can write for the two-flavor neutrino oscillation is \cite{raffeltPRD92}
\begin{equation}\label{mixingHam}
\mathcal{H}=\mix{}_1\aop{}{}^\dagger_1\hat a_1+\mix{}_2\hat a^\dagger_2\hat a_2+\mix{}_m\hat a^\dagger_1\hat a_2+\mix{}_m^*\aop{}^\dagger_2\aop{}_1
\end{equation}
where the prefactors $\mix{}$ are arbitrary complex coefficients. The operator $\aop{}{}^\dagger$ ($\aop{}{}$) represents the creation (destruction) of a particle.

Assuming that particles 1 and 2 correspond to the mass basis, the Hamiltonian becomes
\begin{equation}\label{massHam}
\mathcal{H}_\nu=\sum_{\p{}}\left(\frac{m_1^2}{2p}\aop{}_1^\dagger(\p{})\aop{}_1(\p{})+\frac{m_2^2}{2p}\aop{}_2^\dagger(\p{})\aop{}_2(\p{})\right).
\end{equation}
since the mass states propagate in a given system \cite{mohapatra2004massive}. The energies of the two particles given by $E=\sqrt{p^2+m^2}\approx\mathrm{const.}+m^2/(2p)$.

Had we written this Hamiltonian in the flavor basis, for flavors $\alpha$ and $\beta$, the unitary transformation
\begin{equation}\label{massflavor}
\left(\begin{array}{c}
\aop{}_\alpha(\p{})\\
\aop{}_\beta(\p{})
\end{array}\right)=\left(\begin{array}{cc}
\cos\z{} &\sin\z{}\\
-\sin\z{} &\cos\z{}
\end{array}
\right)\left(\begin{array}{c}
\aop{}_1(\p{})\\
\aop{}_2(\p{})
\end{array}\right)
\end{equation}
can be used where $\theta$ is a mixing angle.  

Note that the term 
\begin{equation}\label{constant}
\sum_{\p{}}\frac{m_1^2+m_2^2}{4p}\left(\aop{}_1^\dagger(\p{})\aop{}_1(\p{})+\aop{}_2^\dagger(\p{})\aop{}_2(\p{})\right)
\end{equation}
is a constant assuming a constant number of neutrinos.  Subtracting it from Eq.~\eqref{massHam} gives
\begin{equation}\label{subtract}
\mathcal{H}_\nu=\sum_{\p{}}\frac\omega2\left(\aop{}_1^\dagger(\p{})\aop{}_1(\p{})-\aop{}_2^\dagger(\p{})\aop{}_2(\p{})\right).
\end{equation}
where
\begin{equation}\label{oscfrequency}
\omega=\delta m^2/(2p)
\end{equation}
and $\delta m^2=m_1^2-m_2^2$.  Isospin operators in the mass basis can be defined based on Eq.~\eqref{subtract} as
\begin{eqnarray}\label{Jcomp1}
{\mathcal{J}}_\p{}^z=\frac12(\aop{}_1^\dagger(\p{})\aop{}_1(\p{})-\aop{}_2^\dagger(\p{})\aop{}_2(\p{})),\\
{\mathcal{J}}_\p{}^+=\aop{}_1^\dagger(\p{})a_2(\p{}),\quad {\mathcal{J}}_\p{}^-=\aop{}_2^\dagger(\p{})a_1(\p{}).\label{Jcomp2}
\end{eqnarray}
which satisfy the SU(2) algebra
\begin{equation}
[{\mathcal{J}}^+_\mathbf{p},{\mathcal{J}}^-_\mathbf{q}]=2\delta_{\mathbf{p}\mathbf{q}}{\mathcal{J}}_\mathbf{p}^z,\quad[{\mathcal{J}}^z_\mathbf{p},{\mathcal{J}}^\pm_\mathbf{q}]=\pm\delta_{\mathbf{p}\mathbf{q}}{\mathcal{J}}_\mathbf{p}^\pm
\end{equation}
Rewriting in terms of the vector $\vec{\mathcal{J}}$, and defining a vector $\mathbf{B}=(0,0,-1)$,
\begin{equation}\label{nu}
\mathcal{H}_\nu=\sum_\mathbf{p}\omega\mathbf{B}\cdot\vec{\mathcal{J}}_\mathbf{p}
\end{equation}
This term resembles the Zeeman term \cite{zeemanN1897,townsend2000modern} which couples an external magnetic field, $\mathbf{B}$, to a spin, $\vec{\mathcal{J}}_\mathbf{p}$.

In the dense neutrino gas, there are also self-refractions of the neutrinos in addition to the flavor oscillations outlined above.  The necessary terms for the Hamiltonian involve four particle terms where $\alpha$ scatters from $\beta$ (also $\alpha=\beta$) \cite{pehlivanPRD11}
\begin{eqnarray}\label{nunu}
\mathcal{H}_{\nu\nu}&=&\frac{\sqrt2G_F}V\sum_{\p{},\q{}}
\vec{\mathcal{J}}_\p{}\cdot\vec{\mathcal{J}}_\q{}
\end{eqnarray}
if the single angle approximation \cite{pehlivanPRD11} is assumed so that scattering is isotropic where $G_F$ is Fermi's constant and $V$ is the quantization volume.  This term provides an interaction between spins. While the above analysis is for states of the neutrino gas, it is supposed that these are the same states available to a single neutrino (see argument in Ref.~\onlinecite{pehlivanPRD11}). Therefore, any analysis of the behavior of the states in the gas corresponds to the states available for the particle.

So far, the analysis has used exclusively the mass basis.  To obtain the suitable form of the unitary transformation for the flavor states instead, we solve the time dependent Schr\"odinger equation \cite{balantekinJPG06}
\begin{equation}
i\frac{\partial U_\mathbf{p}}{\partial t}=\mathcal{H}_\nu U_\mathbf{p}
\end{equation}
where $t$ is the time. The ansatz for $U_\mathbf{p}$ is \cite{balantekinJPG06,pehlivanPRD11}
\begin{equation}
U_\mathbf{p}=e^{\sum_\mathbf{p}z_\mathbf{p}\mathcal{J}^+_\mathbf{p}}e^{\sum_\mathbf{p}\ln(1+|z_\mathbf{p}|^2)\mathcal{J}^z_\mathbf{p}}e^{-\sum_\mathbf{p}z_\mathbf{p}\mathcal{J}^-_\mathbf{p}}
\end{equation}
where $z_\mathbf{p}=e^{i\delta}\tan\theta$.  Performing the unitary rotation ({\it i.e.}, $U^\dagger\hat a_1U$) gives Eq.~\eqref{massflavor}. Defining the flavor isospin operator as $\mathbf{J}_\mathbf{p}=U_\mathbf{p}^\dagger \vec{\mathcal{J}}_\mathbf{p} U_\mathbf{p}$ gives the same functional form for the Hamiltonian, Eqs.~\eqref{nu} and \eqref{nunu}, in flavor space but with $\mathbf{B}=(\sin2\theta,0,-\cos2\theta)_\mathrm{flavor}$.

The form of Eqs.~\eqref{nu} and \eqref{nunu} are identical to Anderson's single particle reformulation of BCS theory \cite{andersonPR58}, which is discussed in Appendix~\ref{andersonBCS}.

To obtain BCS theory, the replacements $\omega\rightarrow2\epsilon_\mathbf{p}$ where $\epsilon_\mathbf{p}$ is the single-particle energy of an electron and of a Cooper pair if $\Delta=0$ ({\it i.e.}, a pair is created and then sent into a region where there is no pair potential),
\begin{equation}
\begin{cases}
J^z_\mathbf{p}=\frac12(a^\dagger_\alpha a_\alpha-a^\dagger_\beta a_\beta)\Rightarrow\frac12(c^\dagger_{k\uparrow}c_{k\uparrow}+c^\dagger_{k\downarrow}c_{k\downarrow})\\
J^-_\mathbf{p}=a^\dagger_\beta a_\alpha\Rightarrow c_{\mathbf{p}\downarrow}c_{\mathbf{-p}\uparrow}\\
J^+_\mathbf{p}=a^\dagger_\alpha a_\beta\Rightarrow c^\dagger_{\mathbf{p}\downarrow}c^\dagger_{\mathbf{-p}\uparrow},
\end{cases}
\end{equation}
for fermionic operators $c$, Eqs.~\eqref{nu} and \eqref{nunu} are the BCS Hamiltonian up to an arbitrary constant.  This provides all necessary information required for the first half of Table~\ref{correspondence} which summarizes the PBKY model.

Arguably the most important quantity in the PBKY model's correspondence (Table~\ref{analog}) is the connection between the paired electron states, captured by the Gor'kov function (see Sec.~\ref{proximity}) \cite{AGD}, and neutrino oscillations.  The connection between neutrinos and Gor'kov function implies that formulating the superconducting case in terms of transport equations for the Green and Gor'kov functions can be used as an analysis tool for neutrino oscillations.

\subsection{Mass and flavor PBKY models as a mean field}\label{meanfield}

Before discussing the transport theory in depth, however, it is useful to show that the external magnetic field derived in the PBKY model is of interest.  Keep in mind that for the condensed matter system, BCS theory is the physical theory and that Anderson's reformulation is an abstraction.  But the neutrino case uses the mass or flavor vector as the physical quantity, describing a neutrino's oscillation, and is real. 

The appearance of a magnetic field on the superconducting state is of importance since the physics of the superconducting state changes greatly if a field is applied versus not applied.  The magnetic field presented so far is not the same as a magnetic field applied on an individual electron.  The PBKY model recognizes the external field as physical for the neutrino oscillations.  For the mass basis, the field provides no source terms.  The flavor basis contains an extra source term and implies the model is written in the mean field limit.

The mass basis Hamiltonian appears as Eqs.~\eqref{nu} and \eqref{nunu}.  This can be simplified to
\begin{eqnarray}\label{genHam}
\mathcal{H}&=&t\sum_\mathbf{p}\omega P^z_\mathbf{p}+H\sum_\mathbf{p}\omega P^x_\mathbf{p}\\
&&+\frac{\sqrt2G_F}V\sum_{\p{},\q{}}P^z_\p{}P^z_\q{}+\frac12(P^+_\p{}P^-_\q{}+P^-_\p{}P^+_\q{}))\nonumber
\end{eqnarray}
where $P$ is either $\mathcal{J}$ (mass) or $J$ (flavor), $P^x=(P^++P^-)/2$, $t=-1$ for mass and $-\cos2\theta$ for flavor, and $H=0$ for mass and $\sin2\theta$ for flavor.  One can view the appearance of a  source term applied for the Gor'kov function in the PBKY model to say that the flavor oscillations themselves are not conserved while the number of neutrinos is allowed to be.

Earlier in writing Eq.~\eqref{constant}, the field was ensured to have a constant number of particles.  This works for the mass basis, but the flavor basis has an external source, $H$, in Eq.~\eqref{genHam} which may not allow for the number conservation on its own.  Thus, this implies that the derived Hamiltonian, specifically the term $P^+_\p{}P^-_\q{}+P^-_\p{}P^+_\q{}$, satisfies the number conservation by explicitly being written in the mean-field of a full term, $P^+_\p{}P^-_\q{}P^-_\p{}P^+_\q{}$ within a constant. Substituting $P^+_\mathbf{p}P^-_\mathbf{q}\rightarrow P^+_\mathbf{p}P^-_\mathbf{q}+\langle P^+_\mathbf{p}P^-_\mathbf{q}\rangle$ returns the mean field for a well-defined constant, $\langle P^+_\mathbf{p}P^-_\mathbf{q}\rangle$.  The connection with BCS theory and the flavor oscillations themselves is more straightforward in this form, especially since interchanging operators to give $-P^+_\p{}P^+_\q{}P^-_\p{}P^-_\q{}$ gives the correct negative sign for the BCS interaction term.  Note that the problem can be phrased in one dimension and the Jordan-Wigner transformation \cite{fetterwalecka} to turn the spin vectors into fermion operators ({\it i.e.}, $P^+_\mathbf{p}\rightarrow c_\mathbf{p}^\dagger$, $P_\mathbf{p}^-\rightarrow c_\mathbf{p}$, and $P_\mathbf{p}^z\rightarrow c^\dagger_\mathbf{p} c_\mathbf{p}-1$) making the connection to BCS theory clear.

The main points of the analysis of the PBKY model is that the Gor'kov function in the BCS superconductivity describes neutrino oscillations and that an external magnetic field is applied for the flavor basis, implying the flavor oscillations themselves have a source term.  This produces very different physics for the BCS ground state than if there were no external field.  The absence of a magnetic field for the mass basis will be recovered below and the external magnetic field will be introduced more easily in scalar field theory and the Standard Model after discussing the transport equations for superconductivity.

\section{Superconducting--magnetic proximity effects}

Transport equations for superconductivity are summarized in Sec.~\ref{proximity} with the resulting physics described in Sec.~\ref{oscillations}.

\subsection{Transport equations for superconducting proximity effects}\label{proximity}

The utility of having identified the PBKY model as equivalent to theories of superconductivity is that this determines the structure of the Green and Gor'kov functions \cite{AGD} as well as the differential operator (from the GL Lagrangian).  From these elements, non-relativistic transport equations can be derived for the superfluid.  A full derivation is contained in Ref.~\onlinecite{bennemann2008superconductivity} (see article by V.~Chandrasekharan), and a summary is provided here.  

\begin{figure}%[b]
\includegraphics[width=\columnwidth]{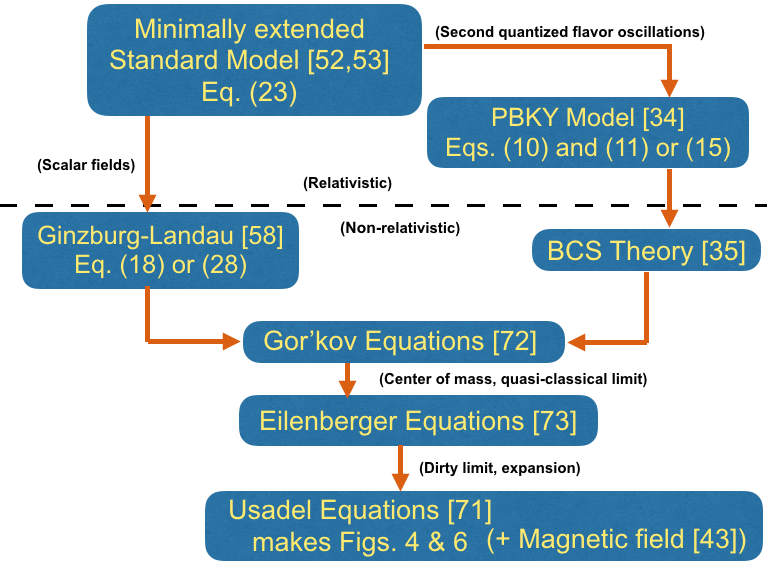}
\caption{The ladder of approximations used throughout this paper is summarized in this flow chart.  The PBKY model is the most fundamental to the arguments presented in the text.  The MESM presented in the text is chosen for its consistency with GL, though it could assume any other form that reduces to GL in the appropriate limit. \label{approxladder}}
\end{figure}

The effect on the superconducting state by the external magnetic field is best exposed from the field theory by following a string of approximations summarized in Fig.~\ref{approxladder}.  The simplifications begin with BCS and GL to reduce the equations and fields to the minimum number of variables needed to describe each Cooper pair.  The resulting Usadel's equations depend only on the center of mass position and energy of each pair \cite{usadelPRL70}.   

Gor'kov connected the microscopic BCS theory and macroscopic GL theory to form what is known as Gor'kov's equations \cite{gorkovJETP59}.  These equations parameterize Cooper pairs in terms of the momentum of each electron, their positions, and their energies.  The quantity of interest is called the Gor'kov function, $\mathcal{F}$, \cite{AGD}
\begin{equation}\label{singlet}
\mathcal{F}=-i\langle0| T\cop{}_{\mathbf{k},\uparrow} \cop{}_{-\mathbf{k},\downarrow}|0\rangle
\end{equation}
with time ordering operator $T$ and $\cop{}$ is a fermionic lowering operator with subscripts for momentum and spin, respectively. Since the superconductor pairs electrons together, it is expected that this correlator creating two paired holes (the time-reversal creates two paired electrons) has a non-zero expectation value in the symmetry broken superconducting phase.  A simplification of Gor'kov's equations by Eilenberger \cite{eilenbergerZP68} rewrites $\mathcal{F}$ in terms of the center of mass momentum, position, and energy.  Further, a fast oscillation in the Green's function is integrated out so the final result is in the quasi-classical limit.  The final simplification from Usadel \cite{usadelPRL70} considers Eilenberger's equations in the diffusive limit where many non-magnetic impurities sufficiently randomize the center of mass momentum on a length scale that is smaller than the characteristic pairing length of the Cooper pair.  This generates a set of equations in the center of mass coordinate and energy for the field as well as the energy.  The key step to arriving at Usadel's equation is to expand the Green and Gor'kov function as L\"uder's has done to give \cite{ludersZNA66,ludersZP68,ludersZP67,luders1971method}
\begin{equation}\label{luders}
\mathcal{F}=f_0+\mathbf{\hat v}\cdot \mathbf{f}+\ldots
\end{equation}
where higher order terms are not necessary, $f_0$ denotes the Gor'kov function for singlet pairing only, $\mathbf{\hat v}$ is the normalized Fermi velocity, $\mathbf{f}$ are the triplet Gor'kov functions, and an average over all momentum is taken as the final step in Usadel's derivation \cite{usadelPRL70,bennemann2008superconductivity}.  Effectively, this expansion expands the field in spherical harmonics to first order \cite{bergeretRMP05}. A similar expression can be found for the clean limit for Eilenberger's equations where one must also include the center of mass momentum of the Cooper pair and an extra matrix structure \cite{eschrigJLTP07}.  For simplicity, the following arguments are constructed in the dirty limit \cite{usadelPRL70}.  Usadel's equations are useful since they often allow for a simple parameterization of the Green and Gor'kov functions and the resulting solution of the parameters clearly displays the behavior of the fully interacting functions, though the same argument can be found from any level of approximation. 

Note that the Green's function, $\mathcal{G}$, of the Cooper pairs also may be expanded as in Eq.~\eqref{luders} \cite{usadelPRL70}.  Since the field $\varphi(x)=\int \mathcal{G}(x,x')q(x')dx'$ in the most general mathematical sense for a forcing function $q$, the expansion of $\mathcal{G}$ shows that the field can be expanded in a similar expansion as Eq.~\eqref{luders} ({\it i.e.}, $\varphi=\varphi_0+\mathbf{\hat v}\cdot\vec\varphi$).  Thus, expanding the mass term ($=-\frac{m^2}2\varphi^2$) in relativistic GL (scalar field) theory to first order with Eq.~\eqref{luders} gives
\begin{eqnarray}\label{GLexpand}
\mathcal{L}^\mathrm{GL}&=&\frac12\varphi\Box\varphi-\frac{m^2}2\Big(\varphi_0^2+\varphi_0(\mathbf{\hat v}\cdot\vec{\varphi})+(\mathbf{\hat v}\cdot\vec{\varphi})\varphi_0\Big)\nonumber\\
&&+\frac{\lambda}{4!}\varphi^4+B\varphi
\end{eqnarray}
where $\Box$ is the D'Alembertian operator and $\lambda$ captures the pairing potential.  The form of the mass term gives the oscillations discussed in this section.

It may be desirable to keep the expansion to second order, but the resulting term is not renormalizable in the particle physics model and would imply extra physics (perhaps another variety of $\varphi_0$ particle).  However, for simplicity, we will limit ourselves to one extra particle.

\subsection{Singlet-triplet oscillations, symmetry protected triplets, and cascading effects}\label{oscillations}

The FFLO effect pairs electrons with momenta $\mathbf{k}+\mathbf{q}$ and $-\mathbf{k}+\mathbf{q}$ instead of $\mathbf{k}$ and $-\mathbf{k}$ as in traditional BCS superconductivity. Two particles (for a simpler example, consider two free particles) with these momenta relations imply an oscillation of the order parameter proportional to $v_F$, the Fermi-velocity \cite{buzdinRMP05}.   The oscillation in the order parameter corresponds to singlet and triplet phases \cite{buzdin1982critical}.

One may describe each pairing state in the $|\ell,s\rangle$ basis, where $s=s_1+s_2$, and note that $\ell=1$ (=0) for the triplet (singlet). Note that applying the interaction term on a singlet pair ($|0,0\rangle$) gives
\begin{eqnarray}\label{tripletcorr}
\mathbf{B}\cdot\mathbf{S}|0,0\rangle&=&(\hat L^-+\hat L^+)(\hat S^z)|0,0\rangle\Rightarrow|1,0\rangle
\end{eqnarray}
since the external magnetic field carries an angular momentum changed by the raising and lowering operators, $\hat L^\pm$.  Note that the spin operator controls which direction the angular momentum is applied.

As pointed out nearly simultaneously by Bergeret-Volkov-Efetov \cite{bergeretPRL01} and Kadigrobov-Shekhter-Jonson \cite{kadigrobovEPL01}, when the angular momentum of a triplet Cooper pair is perpendicular to the applied magnetic field, the triplet components possess $s=\pm1$ and are symmetry protected from the magnetic field.  Perpendicular magnetic configurations are often used in condensed matter systems so that the Cooper pairs may propagate deep into the magnetic material \cite{bakerPRB16}.

So far, the discussion has focused around promoting a singlet pair to a so-called `long-ranged' triplet pair with $s=\pm1$. This is what is generally meant when discussing the (forward) FFLO effect since the $s=\pm1$ pairs can be used to control supercurrents for spintronic application \cite{linderNP15}.  

\begin{figure}[lb]
\includegraphics[width=.6\columnwidth]{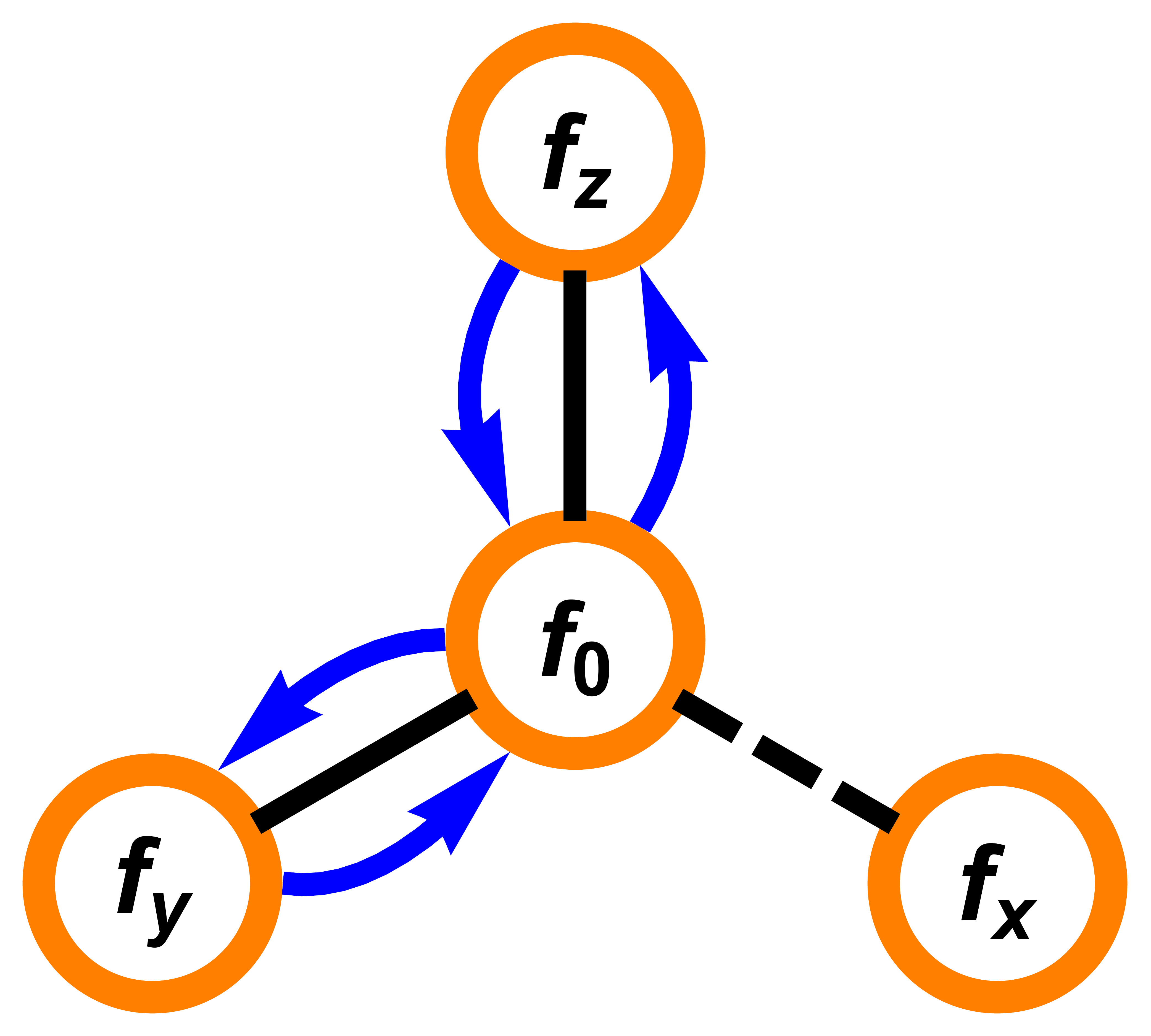}
\caption{A summary of the first order transitions associated with the cascade effect \cite{bakerEPL14} discussion in the text for a Bloch domain wall (two components).  Triplet components may oscillate into the singlet component along the arrows if a field (black lines) are present in that direction.  Broken lines indicate the field never points in that direction.\label{osc}}
\end{figure}

However, the reverse FFLO effects also apply.  For example, applying a field in the direction of an $s=\pm1$ component reverts it back to an $s=0$ component which may undergo singlet-triplet oscillations \cite{bakerEPL14}. In a rotating magnetic field,  the tree-level (see Eq.~\eqref{GLexpand}) transition between a triplet in one of three cartesian directions to a triplet in another cartesian direction produces singlets in the system and can introduce `short-range' ($s=0$) components as though they were `long-ranged' ($s=\pm1$) \cite{bakerNJP14,bakerEPL14,bakerPRB16,bakerJSNM16}.  In general, any direction the magnetic field points can create a triplet component, and this is summarized in Fig.~\ref{osc} for a Bloch domain wall where the external field only points in two directions, generating only two components of $\mathbf{f}$.  This implies that any rotation anywhere in the system will cause a cascade between all available pairing types allowed by the magnetic field \cite{bakerEPL14}.  In general, local and abrupt changes in the magnetization produces high concentrations of $f_0$ components as well as transitions between components of $\mathbf{f}$ \cite{bakerEPL14}.

To first order, this analysis describes what is expected at tree level in Eq.~\eqref{reducedchi} expanded with Eq.~\eqref{neutrinodecomp} and the interpretation of Fig.~\ref{marionette}.  Figure~\ref{osc} may be regarded as the two-flavor version of Fig.~\ref{marionette} (justified in Sec.~\ref{ludersexp}).

\section{Correspondence between neutrinos and superconducting proximity systems}\label{correspondence}

Many statements can be made about the connections made in the previous sections.  The list of topics covered in each subsection are listed here:  sterile and flavored neutrinos from L\"uder's expansion (\ref{ludersexp}), identification of sterile neutrinos as Majorana from symmetry relations (\ref{majorana}), the physics of flavor oscillations by comparison with superconductor-exchange spring systems (\ref{SXSS}), discussion with particular regard to sterile neutrinos in (\ref{f0role}), symmetry protected triplet components as weak process states (\ref{weakstates}), angular momentum in the neutrino case (Sec.~\ref{leptonnum}), MSW effects (\ref{MSW}), CP violation (\ref{CPviolation}), and other possibilities (\ref{exotics}).

\subsection{L\"uder's expansion and neutrino flavor}\label{ludersexp}

The PBKY model establishes that a single neutrino oscillation is equivalent to a Cooper pair represented by a Gor'kov function, $\mathcal{F}$.  To establish each components $f_0,\mathbf{f}$ in Eq.~\eqref{luders} it is first noted that the expansions of Eqs.~\eqref{neutrinodecomp} (see Sec.~\ref{MESM})--which expands in neutrino flavor instead--and \eqref{luders} are in different bases.  Equation~\eqref{neutrinodecomp} expands in flavor or mass while Eq.~\eqref{luders} expands in cartesian directions.  To be more specific, the Eq.~\eqref{luders} expands in the quasi-particle velocity defined generally as \cite{ashcroft1978solidstatephysics}
\begin{equation}\label{velocity}
\mathbf{v}_\p{}\equiv\mathbf{\nabla}_\p{}\epsilon_\p{}=-\frac{\delta m^2}{p^2}\hat p\quad[\mathrm{neutrinos}].
\end{equation}
where the last equality uses the form in the PBKY model for the energy, $\epsilon_{\mathbf{\p{}}}$ ($=2\omega$ for neutrinos outside the gas).  Formally defining this quantity in the diffusive limit where the momentum goes to zero requires the replacement $p\rightarrow p^2+\beta^2$, where $\beta$ is a small parameter on the order of $\delta m$. This allows for an expansion around $p=0$ giving $\hat v=-(\delta m/\beta)^2\hat p$.  In the limit where $\delta m$ and $\beta$, which is seen experimentally for neutrinos, go to zero together, Eq.~\eqref{velocity} is well defined. The oscillation of the superconducting order parameter is proportional to $v_F$, the Fermi-velocity and maximum of $\mathbf{v}_\p{}$, \cite{buzdinRMP05} which is dependent on $\delta m^2$ here.  That Eq.~\eqref{velocity} depends on the mass difference implies that the expansion is for two flavors represented by $\delta m^2$.

  Equation~\eqref{oscfrequency} also implies that Eq.~\eqref{velocity} is related to a similar expression in the mass difference basis.  Connecting Eq.~\eqref{velocity} to Eq.~\eqref{neutrinodecomp} can be accomplished simply by summing Eq.~\eqref{luders} over the three possible mass difference $\delta m^2$.  The coordinate system for the problem is also altered to one where the neutrino's momentum is always pointing in one direction for simplicity (this is satisfied if the neutrino is restricted to one dimension, for example).  Another route to changing the basis is to use Eq.~\eqref{osc} for each of the three mass differences and corresponding momenta.
  
Note that summing Eq.~\eqref{luders} seems to imply the need for one $f_0$ (equivalently $\chi_0$) component for each generation in $\mathcal{L}_\sm{}$ as was mentioned earlier in a different context in Sec.~\ref{MESM}. The present analysis will leave the $f_0$ ($\chi_0$) component as a single field. Summing over mass differences allows for one to replace Eq.~\eqref{luders} by Eq.~\eqref{neutrinodecomp}, effectively expanding the field in flavor going forward.

Repeating the derivation of Usadel's equation for an expansion in mass differences has no effect on the final form of Usadel's equations \cite{bennemann2008superconductivity}.  So, the vector components, $\mathbf{f}$, can be regarded as representing the three neutrino flavors up to a unitary transformation.

%Using either representation, Eq.~\eqref{neutrinodecomp} or Eq.~\eqref{luders}, leads to the same conclusions in Sec.~\ref{SXSS} and Sec.~\ref{f0role}, but the expansion in flavor is more straightforward.

Having identified the correspondence between the vector $\mathbf{f}$ and the three flavors of neutrino, the component $f_0$ is a scalar in the condensed matter case and is invariant under rotation of the applied field.  Contrastingly, $\mathbf{f}$ transforms like a pseudo-vector since it has acquired an angular momentum.  This invariance of the singlet already points to its correspondence to the sterile neutrino, which is expected to not carry a flavor.

\subsection{Quasi-particle symmetry relations and Majorana fermions}\label{majorana}

The time reversal symmetry relations of the components in the condensed matter system are \cite{eschrigJLTP07}
\begin{eqnarray}\label{symmetries}
\tilde{f}_0(\omega_n)=f_0(-\omega_n)\quad\mathrm{and}\quad\mathbf{\tilde{f}}(\omega_n)=-\mathbf{f}(-\omega_n)
\end{eqnarray}
at each position where $\omega_n$ is the Matsubara frequency \cite{AGD}.  This reflects the nature of the triplet pairing to be odd in frequency but even in momentum (the limit $\p{}\rightarrow0$ for the diffusive limit is shown) under time reversal denoted by a tilde.  Note that these symmetries can be derived from the transport equations by examining the symmetry of the pairing potential, $\Delta(\mathbf{r})$ \cite{eschrigJLTP07}.  The symmetry relations do not depend on the energy scale or the spin, so the reduction of Eq.~\eqref{minimalSM} to Eq.~\eqref{reducedchi} should not cause any issues in using Eq.~\eqref{symmetries}.

The time reversal symmetry operator in the quantum field theory, based on the symmetries of the Dirac equation, does not mix particle and anti-particle components of the 4-vector \cite{peskin1995introduction}.  This implies that Eq.~\eqref{symmetries}, for the singlet $f_0$ cannot be a Dirac particle since it relates the particle (pairs with $\omega>0$ as assigned in the PBKY model) with the antiparticle ($\omega<0$).  The neutrino is also not likely to be a scalar particle since it is expected to have non-zero spin.  The remaining possibility is that the neutrino is a Majorana particle \cite{wilczekNP09}.  The Majorana particle has no distinction between particle and anti-particle as they are equivalent, so the Majorana particle satisfies the symmetry relations given.   The field corresponding to $\mathbf{f}$ can be a Dirac field \cite{dodelsonPRL94}.

For the condensed matter case, note that the Cooper pair is a Majorana particle when the weights of the Bogolyubov-Valatin transformation \cite{valatinINC58,bogolubovJP66} are equal which is physically realized in $p+ip$ superconductors \cite{wilczekNP09}.  The formalism used, that of Usadel, is still valid for these systems \cite{tanakaPRL07}.

Another type of Majorana mode may be found in magnetic systems with a superconducting gap  \cite{fuPRL08} but this is a distinct type of Majorana mode from the fermion of present interest.  The individual Majorana particle is different from the topological state of matter with regards to the properties of the sterile, $f_0$ neutrino.  The superconducting gap is realized in the neutrino gas, so the topological Majorana mode may be realized in the neutrino gas's case.  In any case, this is distinct from the Majorana fermion.

\subsection{Minimally extended Standard Model}\label{MESM}

Given that BCS theory appeared for neutrino oscillations in the PBKY model, the Lagrangian formulation is provided by GL theory.  This sections shows that a reduction of the MESM reduces to the GL Lagrangian, although it is possible that other model extensions also reduce to the desired result.  This is the simplest extension that does not change the symmetry group of the model.  The appearance of the sterile neutrino is justified by its appearance in the superconducting case (see Sections~\ref{ludersexp} and \ref{majorana}).

The simplest term one can add to the Standard Model to obtain Eq.~\eqref{mixingHam} is the Pontecorvo-Maki-Nakagawa-Sakata (PMNS) matrix \cite{pontecorvoJETP68,makiPTP62}. Incorporating a Majorana mass term  that couples left-handed particles together is an addition to the Standard Model, $\mathcal{L}_\sm{}$, that appears as \cite{drewesIJMPE13}
\begin{equation}\label{leftleft}
\mathcal{L}=\mathcal{L}_\sm{}-\frac12\bar\nu_Lm_\nu\nu_L^c+\mathrm{h.c.}
\end{equation}
where $\nu_L$ is a left-handed Majorana field with mass $m_\nu$ for each neutrino and $c$ is the charge conjugation operator. When this term is written for energies above the electro-weak symmetry breaking, Eq.~\eqref{leftleft}--requiring two Higgs bosons--is of mass dimension 5 \cite{dodelsonPRL94,peskin1995introduction} which requires a coupling constant of negative mass dimension.  Such a term is therefore non-renormalizable and indicates that some degree of freedom has been integrated out \cite{schwartz2013quantum}. 

A candidate for the integrated out quantity is a right handed sterile neutrino.  The Standard Model can be extended to include a right handed field for the sterile neutrino, $\nu_0$, \cite{drewesIJMPE13}
\begin{eqnarray}\label{minimalSM}
\mathcal{L}&=&\mathcal{L}_\sm{}+i\bar{\nu}_0\slashed{\partial}\nu_0-\bar{\mathbf{E}}_LF\nu_0\tilde{\Phi}-\bar{\nu}_0F^\dagger\mathbf{E}_L\tilde{\Phi}^\dagger\\
&&-\frac12\left(\bar{\nu}_0^cM_M\nu_0+\bar{\nu}_0M_M\nu_0^c\right)\nonumber
\end{eqnarray}
where $F$ is a tensor of Yukawa couplings, $\Phi$ is the Higgs boson with $\tilde\Phi=(\epsilon\Phi)^\dagger$ where $\epsilon$ is the SU(2) anti-symmetric tensor, and $\mathbf{E}_L$ is a vector of doublets for each generation in $\mathcal{L}_\sm{}$.  The possibility of several sterile neutrinos are allowed with the Majorana mass matrix, $M_M$.  Equation~\eqref{minimalSM} contains both the Majorana and Dirac masses.  Only one sterile neutrino is required based on arguments in Sec.~\ref{proximity} but more sterile neutrinos may be desirable \cite{drewesIJMPE13,canettiPRL13}.  The following arguments are without a loss of generality to more sterile neutrinos and $M_M$ is reduced to scalar $m_0$ in the following.

Taking Eq.~\eqref{minimalSM} at low energies and in the case of free neutrinos, the equations of motion with respect to $\bar\nu_\rho$ and $\bar\nu_0$, respectively, can be written as
\begin{eqnarray}\label{EOM1}
i\slashed{\partial}\nu_\rho&=&m_\rho\nu_0\\
i\slashed{\partial}\nu_0&=&\frac{m_0}2\nu_0^c+\sum_\rho m_\rho\nu_\rho\label{EOM2}
\end{eqnarray}
where an index $\rho$ indexes the neutrino flavors, $\rho\in\{e,\tau,\mu\}$.  Adding Eqs.~(\ref{EOM1}--\ref{EOM2}),
\begin{equation}
i\slashed{\partial}\nu_0+\sum_\rho i\slashed{\partial}\nu_\rho=\frac{m_0}2\nu_0^c+\sum_\rho m_\rho\nu_0+m_\rho\nu_\rho.
\end{equation}
The constants multiplying each field are of the same order of magnitude, and so an approximation that all magnitudes, $m$, are equal is made for simplicity.  

Converting all fields to a scalar field (as is done commonly in scalar quantum electrodynamics calculations \cite{schwartz2013quantum} since the neutrino's spin should not affect flavor oscillations), the equations of motion can be rewritten in terms of one master scalar field,
\begin{equation}\label{neutrinodecomp}
\chi=\chi_0+\mathbf{\hat{w}}\cdot\mathbf{\vec\chi}
\end{equation}
with $\vec\chi=\langle\chi_e,\chi_\tau,\chi_\mu\rangle$ and $\mathbf{\hat w}$ represents the flavor basis unit vector.  This expansion is reminiscent of the symmetry breaking expansion used in the linear sigma model (but is used in the same way as the condensed matter system in Sec.~\ref{ludersexp}).

The Lagrangian in the scalar field $\chi$ is
\begin{equation}\label{reducedchi}
\mathcal{L}(\partial\chi,\chi)=\frac12\chi\Box{}\chi-\frac{m^2}2\chi^2+B\chi
\end{equation}
where $B\chi$(=$B\chi_0+B\sum_\rho\chi_\rho$ from Eq.~\eqref{neutrinodecomp}) is introduced in a purely mathematical sense for the purpose of calculating correlation functions.  The physical origin of this source field may come from several places.  Terms that were neglected in the analysis so far ({\it i.e.}, coupling of flavored neutrinos to leptons via gauge bosons--in particular the $W$ boson) might account for the external field coupling to the flavored states ($B\sum_\rho\chi_\rho$).  The coupling of the field to the sterile state ($B\chi_0$) may come from many sources \cite{konetschnyPLB77,gelminiPLB81,chenPRD07} or is an artifact of writing the neutrino fields as one field.  The physical origin of this field is not of particular interest here and it is sufficient to introduce it mathematically.

Note that one term is missing from Eq.~\eqref{reducedchi} that appears in GL, a density-density term proportional to $\chi^4$ must be added for the neutrino gas (this term is equivalent to Eq.~\eqref{nunu}).

The field $\chi$ represents a neutrino particle whose oscillations are different representations of $\chi$ in the expansion of Eq.~\eqref{neutrinodecomp}. There is a question of momentum and energy conservation for such a form:  since the mass of the three neutrino flavors are different, a free neutrino can not conserve its momentum and energy while changing its mass \cite{cohenPLB09}.  This point will be revisited in Sec.~\ref{weakstates} when it is shown in the proximity system that turning off the external field freezes the oscillations between the components.

\subsection{Flavor and mass basis: Connection to superconductor-exchange spring proximity systems}\label{SXSS}

Neutrinos are typically written as in Eq.~\eqref{massflavor} where it is recognized that the particle has two different representations:  the neutrinos can either be written in terms of the mass basis or the flavor basis.  The classic description of mass-flavor oscillations is that a propagating neutrino will oscillate flavors by first oscillating into one of the mass states and then back to a flavor state which may not be the same as the original flavor \cite{pehlivanPRD11}.  This rotation in mass-flavor space is connected rigorously in the PBKY model by a unitary transformation between the fields describing the mass and flavor states ({\it i.e.} rotating the fields in Eqs.~(\ref{Jcomp1}--\ref{Jcomp2})).

However, the superconducting-magnetic system does not use this two field (mass and flavor) construction.  Instead, one can observe that a unitary rotation of the Hamiltonian Eq.~\eqref{nunu} and \eqref{nu} can equivalently be phrased in terms of a rotating magnetic field ($\mathbf{B}$) instead of rotating fields ($\mathbf{J}$). Thus, the description of the neutrino oscillations will only involve one basis:  the mass-flavor basis with a rotating preferred direction provided by the sources of the field.  

The two-flavor case corresponds to a rotating Bloch domain wall as in the superconductor-exchange spring system of Ref.~\onlinecite{bakerNJP14} which contains a graphical depiction of the system.  Only two of the three cartesian directions have a non-zero magnetization in the Bloch domain wall, giving non-zero amplitude to two of the components of $\mathbf{f}$, and this makes the Bloch domain wall an ideal system to study the two-flavor physics of the PBKY model.

In the gapless case (free neutrinos), the Cooper pair will eventually break at some distance from the superconductor, and the physical effects of interest, namely the FFLO effect, appears close to the superconductor, typically within a few nanometers, though that length scale may be much different in the neutrino case as the diffusion coefficient is defined differently.  The gapless case is the vacuum limit for the neutrinos and this pair breaking can be regarded as a loss of quantum coherence from the source if the neutrino is a fundamental particle, though it is not explicitly ruled out here that it may be composed of other particles.

The complete computational details can be found in Ref.~\onlinecite{bakerPRB16} where a parameterization by Ivanov-Fominov \cite{ivanovPRB06,ivanovPRB09} which parameterizes the Green and Gor'kov functions in terms of trigonometric functions \footnote{The normalization condition in Ref.~\onlinecite{ivanovPRB06}, $M_0^2-|\mathbf{M}|^2=1$, reminds of a relativistic invariant.  The condensed matter system is represented as a SU(2)$\otimes$SU(2) theory and is a double cover of the Lorentz symmetry, so relativistic invariants are not wholly surprising.}. This parameterization allows for the exact identification of singlet and triplet pairing.  It is useful to examine the system in one dimension where the physics of the full three dimensional system can be exposed with a reduced computational cost with reliable results in comparison with experiment.

Having identified all the components of $\mathcal{F}$ in the condensed matter case with their corresponding particle physics representation in the previous section, now the results of previous works \cite{bakerNJP14,bakerEPL14,bakerPRB16} are used to implement a solution of Usadel's equations for a superconductor-exchange spring system and identify behaviors of the neutrino oscillations.

\begin{figure}
\includegraphics[width=\columnwidth]{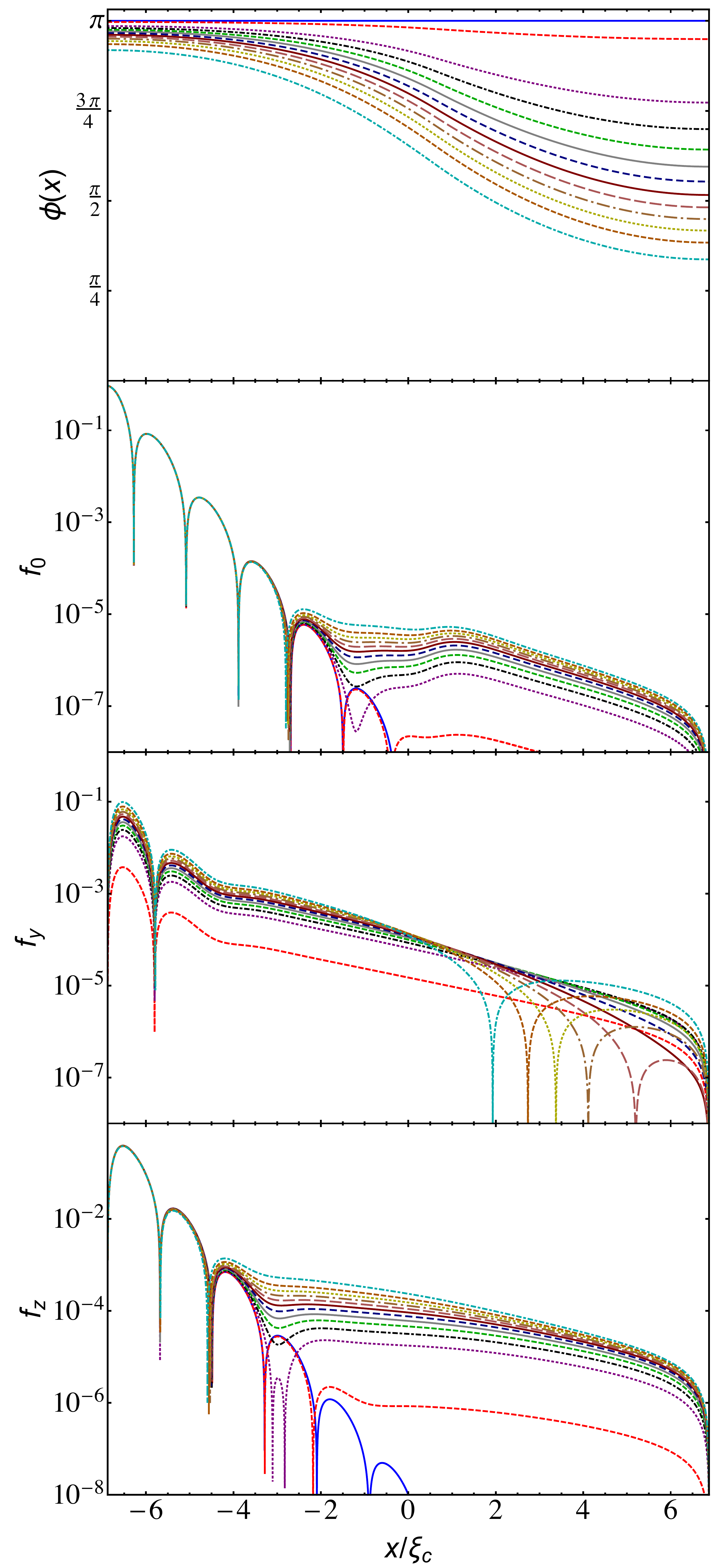}
\caption{The domain walls, singlet, and triplet Gor'kov functions.  See text for a full discussion. Parameters for the ferromagnetic system are chosen to be most closely aligned with a Cobalt/Permalloy system with parameters $K_1/K_2=625$, $A_1/A_2=1$, domain walls found at $h=14\pi T_c$ for Co and $8\pi T_c$ for Py, with an interface placed at $x=0.625\xi_c$, a total distance of $d_F=13.75\xi_c$, $T=0.2T_c$, and $\Delta=0$.\label{ungapped}}
\end{figure}

\begin{figure}[b]
\includegraphics[width=.8\columnwidth]{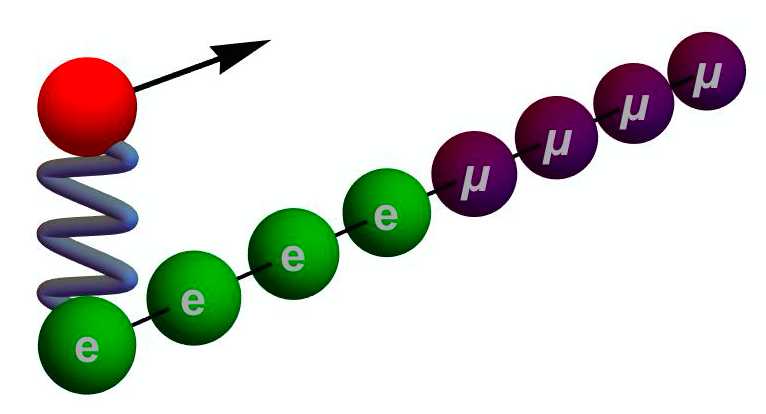}
\caption{A proximity system with two homogeneous ferromagnets oriented perpendicularly to each other \cite{bakerPRB16} corresponds to the case where the sterile neutrino, $\nu_0$, encounters a background of one type of lepton and then another.  In this depiction, it encounters a background of electron lepton flavor and then the $\mu$ lepton flavor.  This is a rotation in isospin space of the external field (wavy line) is analogous to the rotating the magnetic field in cartesian coordinates in a magnetic system consisting of two perpendicular ferromagnets.
\label{cartoon2}}
\end{figure}

\subsection{Role of $f_0,\chi_0$ in proximity systems and flavor oscillations}\label{f0role}

Shown are four panels in Fig.~\ref{ungapped} corresponding to the domain walls of the twisted exchange spring corresponding to the rotation of the domain wall (top panel), the singlet pairing (second panel), triplet pairing in the $\mathbf{\hat y}$ direction (third panel), and the triplet pairing in the $\mathbf{\hat z}$ direction (last panel). One may think of the superconductor (not pictured; left of amplitudes) as an arbitrary bath for neutrinos corresponding to the $f_0$ component (sterile neutrino) in the system characterized by $f_0(\omega_n)=\Delta/\sqrt{\omega_n^2+|\Delta|^2}$ for some pairing potential, $\Delta$, and $\mathbf{f}=0$.   All components are set to zero on the opposite side \cite{bakerPRB16}. Singlet and triplet amplitudes oscillate when a magnetic field is oriented in a constant direction (shown by oscillations on the figures) and correspond to $s=0$.  Meanwhile, the symmetry protected states ($s=\pm1$) decay exponentially with a longer coherence length--a behavior characteristic of normal metals (no magnetic field). The long range decays (perpendicular components of the Gor'kov function to the field) can be assigned to mass eigenstates and oscillating curves (parallel components) to flavor eigenstates \cite{bakerPRB16}, following Eq.~\eqref{genHam}.

As the magnetization twists more and more in the top panel of Fig.~\ref{ungapped} (descending curves), more of the $f_0$ states are deposited into the system as in Fig.~\ref{osc}. The singlet/sterile states appear to decay exponentially without oscillation in the interval $x>-2\xi_c$.  These `short-ranged' components appear more as a `long-ranged' triplet component with an exponential decay, albeit a orders of magnitude lower than the triplet states.  In truth, these exponentially decaying singlets were present even close to the superconductor but only reveal themselves with the oscillating singlets from the superconductor ($x<-2\xi_c$) decay enough in their amplitude to reveal the exponential behavior \cite{bergeretPRB01c,bakerPRB16,bakerJSNM16,bakerJMMM16}.

Reference~\onlinecite{mentionPRD11} suggests that sterile neutrinos can be found close to sources.  This is wholly consistent with this analogy.  In that situation, the flavored neutrinos would be produced and oscillate into the sterile state so long as an external field is present to change the flavor.  That the sterile neutrinos are observable near the source merely implies that there are abrupt flavor oscillations near this region.  Away from this region, the flavor oscillations do not interact with the field and produce less sterile components.  Note also that $f_0$ need not be large even in a slow but constant rotation of the field.  Contrastingly, abrupt rotations of the domain wall cause large populations of singlets to appear, for example, in discrete rotation of the domain wall \cite{bakerEPL14,bakerPRB16,bakerJSNM16,bakerJMMM16}.  The analogous situation for neutrino oscillations is where the conserved background current is abruptly changed as is depicted in Fig.~\ref{cartoon2}.

The bottom two panels in Fig.~\ref{ungapped} correspond to components of $\mathbf{f}$.  The component $f_x$ is zero everywhere since the magnetization never points in this direction.  Again, oscillations corresponding to $s=0$ components are seen, and so are $s=\pm1$ components with exponential decays propagating much further into the system.  Note that expressing these components as parallel and perpendicular  to the field shows clear `short' and `long' ranged behavior and is discussed in Ref.~\onlinecite{bakerPRB16}.

\begin{figure}
\includegraphics[width=\columnwidth]{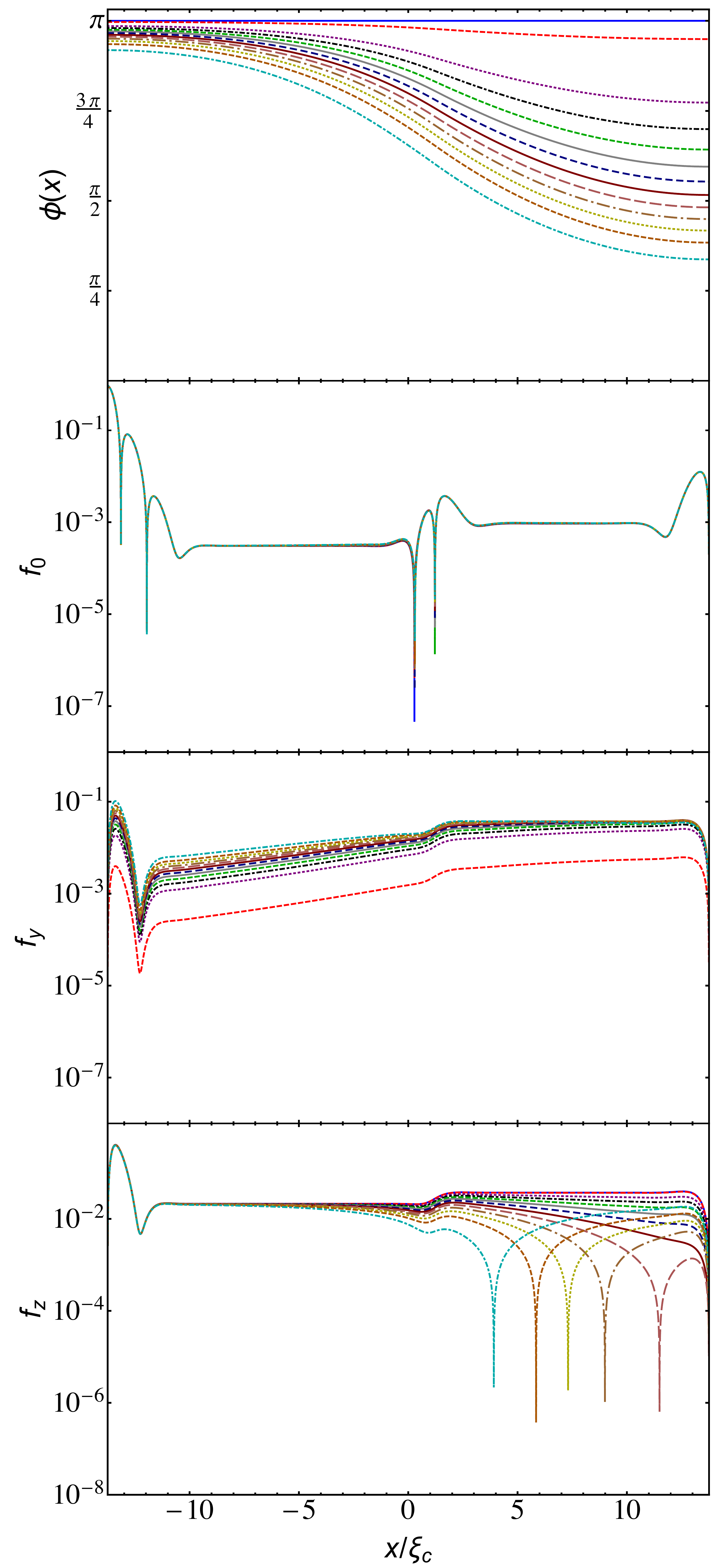}
\caption{The same as Fig.~\ref{ungapped} but with $\Delta=0.3\pi T_c$ and $d_F=27.5\xi_c$.  The same angles at the edges of the ferromagnet were used as in Fig.~\ref{ungapped} for comparison purposes only and not recalculated for the longer layer; a longer layer is called for to avoid a finite size effect for this gapped case; this can be thought of as changing the domain wall length or reducing the coherence length $\xi_c=\sqrt{D/(2\pi Tc)}$ for some diffusion coefficient, $D$, and critical temperature $T_c$.  Effectively, the domain wall from Fig.~\ref{gapped} is re-scaled for this longer system.\label{gapped}}
\end{figure}

Adding back in the pairing amplitude of the superconductor corresponds directly to the neutrino gas.  This allows for neutrino-neutrino scattering in Eq.~\eqref{nunu} and is represented by the BCS order parameter.  Fig.~\ref{gapped} shows the superconductor-exchange spring with a gap for reference. The oscillations are evident near the bath of $f_0$ but neutrino amplitudes saturate far from the source to a constant value.  The domain walls of the exchange spring are chosen to match Fig.~\ref{ungapped} even though the distance has been increased by a factor of two.  This is to avoid a finite size effect from the extremely long correlation length of the pairing potential, $\Delta$.  Saturation to two different values on either side of the interface ($d_F=0.625\xi_c$ ungapped; $d_F=1.25\xi_c$ gapped) reflects the changing magnetization strength.  There is a jump in the value at the interface also for these reasons.  As one twists the domain wall, the singlet is relatively unchanged due to its rotational invariance.

The singlet in Fig.~\ref{gapped} is at least an order of magnitude lower than the triplet amplitudes which also saturate.  Oscillations can be seen to the right of the interface in the $f_z$ component as the domain walls increase their twist and more singlets are generated on that half of the magnetic system.

The summary statement is that the sterile neutrino controls the oscillations of the flavored neutrino states.  The amplitude of the sterile neutrino is much less than the flavored states and is directly controlled by the magnitude and rotation of the applied field.

\subsection{Weak process states and symmetry protected triplet components}\label{weakstates}

Triplet states with $s=\pm1$ have the same character as a weak process state in the neutrino case \cite{giuntiPRD92,kruppke2007theories,cohenPLB09}.  Flavored neutrino states require modification so that decays involving the weak force guarantee the creation of the correct flavor ({\it i.e.}, beta decays must produce electron-type anti-neutrinos where the guarantee of the electron type is desired).  

Reference~\onlinecite{cohenPLB09} uses this idea to construct a theory where beta decays are guaranteed to give electrons and electron-type neutrinos only but also that entangle leptons and their partner neutrinos.  After some distance, Ref.~\onlinecite{cohenPLB09} presumes that these entangled states decay and the projection of the neutrino's flavor onto the other isospin directions gives the oscillation.  According to the PBKY model and the discussion above, these protected states do exist in analog with the $s\neq0$ triplets components when there is no external field of like flavor present, but upon encountering another region where the external field points in the same direction as the triplet's angular momentum, the symmetry protected state reverts to a $s=0$ triplet and may undergo oscillations. In other words, if the isospin of the symmetry protected state does not match the isospin carried by the external field, then there is no reversion to the $s=0$ triplet states and the state remains symmetry protected. Reference~\onlinecite{cohenPLB09} might be regarded as an effective theory of only the symmetry protected states but not of the oscillations themselves.  

With regards to the energy and momentum conservation discussed after Eq.~\eqref{reducedchi}, the situation can be resolved if the neutrino does not oscillate (mass does not change).  According to the condensed matter system, this occurs when there is no external field or a field that is perpendicular to the flavor direction of the neutrino.  This was the physics derived in Eq.~\eqref{genHam} in the PBKY model where there was no field found for the mass basis.

\subsection{Angular momentum and lepton number}\label{leptonnum}

The clearest meaning of $\ell$ available from the PBKY connection is replacing the moment of inertia by a quantity in flavor space to give a flavor angular momentum.

It is tempting to call the angular momentum quantum number, $\ell$, for the neutrinos as a lepton number.  This is since the sterile state has $\ell=0$ and the flavor components that have $\ell=1$.  However, it is not clear if the quantum number appears similarly for leptons in the theory.  Note that if components of $\mathbf{f}$ leave the region where the external field may act on it, it remains in the symmetry protected state (symmetry protected states from a ferromagnet survive in a normal metal \cite{cottetPRL11}).  So, $\ell$ will not change unless in response to an external field.
 
\subsection{FFLO and MSW effects}\label{MSW}

Taking into account the effects of the external field from Sec.~\ref{SXSS}, the behavior of the components of $\mathbf{f}$ behave similarly to neutrinos in the presence of matter.  The MSW effect is a modification of neutrino couplings or masses in the presence of matter \cite{wolfensteinPRD78,mikheyevINCC86,mohapatra2004massive,pal2014introductory}.  This can cause mixing angles \cite{olive2014review} to alter.  These effects are observed, for example, inside of the sun where the high density of leptons alters these couplings. There is also experimental evidence that neutrino fluxes are greater when viewing them coming from the earth as opposed to from space \cite{fukudaPLB94}.  Both of these effects have a similarity with the proximity effects in that they are polarizing the flux of particles moving through the system.  If only electrons are around, then it is expected that more electron type neutrinos and the same effect occurs if the external field either has a larger magnitude or points in one direction in a region of space.  In the condensed matter system, if the magnetic field is oriented in the $\mathbf{\hat z}$ direction, more triplets in the $\mathbf{\hat z}$ direction appear. This is the essence of the MSW effect.

That the MSW effect appears so similar to a particular orientation of the domain wall implies the external field is connected to leptons, but this can be no more than a conjecture based on the analysis here.

\subsection{FFLO phases and CP violation}\label{CPviolation}

Taking the suggestion that an external field splits the momentum the Cooper pair, written $u_{\mathbf{k}}{}c_{\mathbf{k},\uparrow}+v_{\mathbf{k}}c_{-\mathbf{k},\downarrow}^\dagger$ with coefficients  $u,v$ \cite{valatinINC58,bogolubovJP66}, the same relation can be applied to a Majorana spinor in its quantized form as \cite{peskin1995introduction}
\begin{eqnarray}\label{majpart}
\chi(\p{})=\left(\xi_{-\p{}}\hat a_{-\p{}}+(i\sigma_y)\xi_\p{}^*\hat a_\p{}^\dagger\right)\delta(p^2-m^2)
\end{eqnarray}
for a spinor $\xi$.  Under the FFLO effect, having a momentum $\mathbf{q}$ being transferred by the external field, this becomes
\begin{eqnarray}\label{majsplit}
\chi(\p{},\mathbf{q})&=&\xi_{-\p{}+\mathbf{q}}\hat a_{-\p{}+\mathbf{q}}\delta((-p+q)^2-m^2)\\
&&+(i\sigma_y)\xi_{\p{}+\mathbf{q}}^*\hat a_{\p{}+\mathbf{q}}^\dagger\delta((p+q)^2-m^2).\nonumber
\end{eqnarray}
This breaks the symmetry between the parts of the Majorana field that are particles and those components that are anti-particles in a way that is similar to the splitting of paired electrons.  This would add an extra phase factor to calculated scattering amplitudes.  A weak field produces a weak splitting.

In the CP violation problem, a phase factor can account for particle-anti-particle a-symmetry seen in the universe \cite{doiPLB81,wolfensteinPLB81}.  The entire discussion of Majorana particles that are split by a complex phase as in Eq.~\eqref{majsplit}, which in effect adds a factor of $e^{i\mathbf{q}\cdot\mathbf{x}}$ to the Majorana, is very reminiscent of this issue.  The tangible consquence is that when calculating a given scattering amplitude with the neutrino in the presence of an external field, the extra phase appears.   This phase produces complex phase factors in the PMNS matrix and Cabibbo-Kobayashi-Maskawa (CKM) \cite{cabibboPRL63,kobayashiPTP73}.

Experimentally, the violation can be measured in kaon oscillations of color singlets (consisting of, for example, a strange and down quark) as moderated by the external fields \cite{peskin1995introduction}.  It is conjectured that this also occurs for leptons.  The other place where it is expected that the complex phase to enter is in the derived mass term above.  The neutrinos acquire this phase in the $m\bar{\nu}_0\nu_\rho\rightarrow e^{i\mathbf{k}\cdot\mathbf{x}}m\bar{\nu}_0\nu_\rho$ where $\mathbf{k}$ contains the momentum of the sterile and flavored neutrino. Note that the phase factor considered here is different from the Majorana phase (set to one throughout this paper) \cite{mohapatra2004massive,pal2014introductory}.

For other particles that do not undergo flavor oscillations, there is no equivalent external field in the PBKY model ({\it i.e.}, Eq.~\eqref{nu} does not appear since Eq.~\eqref{mixingHam} does not mix 1 and 2).  This is one reason why leptons may not undergo this effect, though it is not ruled out. Two Dirac particles that are paired together, however, as in the kaon's or Cooper pair's case, form an effective particle that can receive a phase by the FFLO effect, for example.

Many other explanations may account for CP violation \cite{pecceiPRL77,banerjeePLB03}.

\subsection{Further consequences and exotic possibilities}\label{exotics}

Some exotic possibilities may also be realized.  The literal analog of the connection between Josephson junctions and neutrinos is the appearance of a current flowing between two superconductors in the former and the transfer of particles between two interstellar bodies in the latter, but the same mechanism may appear in both systems (though the stellar phenomenon might be more of an instantaneous transfer akin to a lightning bolt as a local phase may be responsible in the particle physics case).  

Many non-intuitive effects exist in the proximity effect literature.  It may be possible to realize these in the context of the particle or astrophysics sense such as superharmonicity in Refs.~\onlinecite{trifunovicPRL11} and \onlinecite{richardPRL13}.

The single angle approximation from Ref.~\onlinecite{pehlivanPRD11} restricts the equivalent superconductor to that of an $s$-wave \cite{bennemann2011physics}.  If there is a situation where this approximation breaks down, other types of pairing symmetry may be realized in the neutrino gas, such as $p$-wave.

The general discussion of FFLO effects in this article applies to any flavor mixing process corresponding to Eq.~\eqref{mixingHam}.  For example, FFLO effects on bound quark pairs describe glitches in the radio frequency pulses of a neutron star \cite{rajagopalPRL01}.  The singlet and triplet states used in this article are most likely not significant for bound quarks (such as in color-superconductivity) in the same way as for neutrinos.  For the quark case, the paired quark states correspond to a single Cooper pair. The CKM matrix, which is the analog of the PMNS matrix, is therefore describing two particles bound together which is different from the analysis in this paper ({\it i.e.}, a mixing matrix describing one particle with the FFLO effect). 

\section{Conclusion}

This paper connects the physics of neutrino flavor oscillations and superconducting proximity effects.  The FFLO effect causes singlet-triplet oscillations of the superconducting state that physically resemble flavor oscillations in the neutrino case.  This demonstrates the physical need for sterile neutrinos since they correspond to singlet states in the superconductor.  Both are necessary for triplet/flavored states to oscillate.

Neutrino oscillations can either be symmetry protected (analogous to mass eigenstates) or oscillate into another neutrino type (flavor eigenstates).  This allows for the conservation of mass and energy in oscillations by only allowing them when an external source field is present.  A cascade between all neutrino types available in a given system occurs when a changing external field is present.

The symmetry relations of the quasi-particle states in the condensed matter case show that the sterile neutrino is a Majorana fermion.  The sterile neutrinos are the intermediate states between other flavor states and appear wherever neutrinos are found--especially when the external field is rotated abruptly through the direction of the neutrino flavor. If the sterile neutrino is set to zero, no flavor oscillations occur.

A quantum number analogous to the angular momentum quantity of the proximity system is defined for a neutrino; the sterile state has a quantum number of zero while the flavored neutrinos have a value of one.  Weak process states bear a close similarity with symmetry protected $s=\pm1$ triplet states of the superconductor.

Other effects such as the MSW effect are contained in the physics of the proximity effect since both polarize the field. Phase factors acquired from the FFLO effect might account for the CP violation seen in kaon oscillations by adding the appropriate phase factors and also allow for their effects in neutrinos.

\section{Acknowledgements}

Support is gratefully provided by the Pat Beckman Memorial Scholarship from the Orange County Chapter of the Achievement Rewards for College Scientists Foundation.

\begin{appendix}

\section{Overview of Anderson's reformulation of BCS superconductivity}\label{andersonBCS}

In the rewritten theory of Ref.~\onlinecite{andersonPR58}, the vectors $\mathbf{J}_\mathbf{p}$ (or $\vec{\mathcal{J}}_\mathbf{p}$) appear as spin vectors. In a slight departure from the PBKY model, the Hamiltonian appears as \cite{andersonPR58}
\begin{equation}
\mathcal{H}=\sum_\mathbf{k}\mathbf{H}_\mathbf{k}\cdot\mathbf{J}_\mathbf{k}
\end{equation}
where $\mathbf{H}=2(\epsilon_k-\epsilon_F)\mathbf{\hat z}+\sum_\mathbf{k'}V_{kk'}\mathbf{J}_{\perp\mathbf{k'}}$ with $\mathbf{\hat z}$ picking the $z$ component of $\mathbf{J}_\mathbf{p}$ and ``$\perp$" selecting the perpendicular components $x$ and $y$.  Note that a Fermi-energy, $\epsilon_F$, was defined and is equal to the chemical potential at zero temperature.

\begin{figure}[b]
\includegraphics[width=.8\columnwidth]{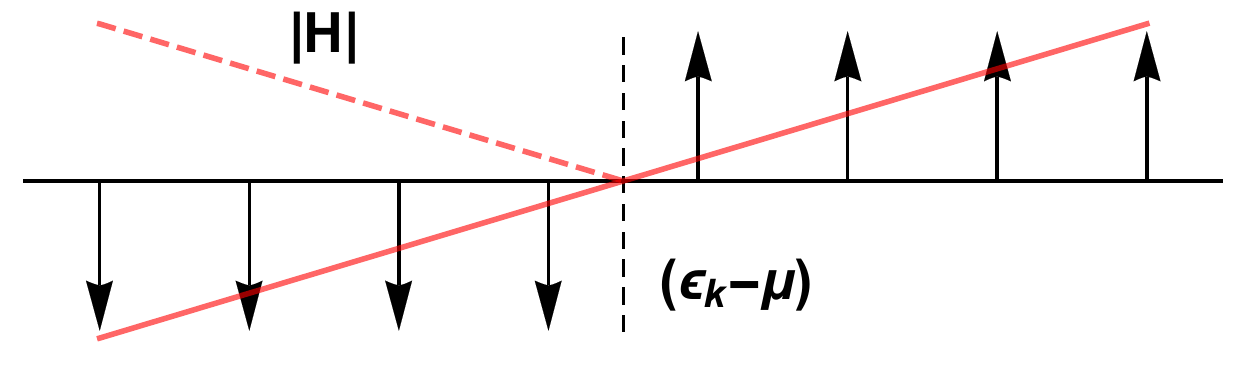}
\includegraphics[width=.8\columnwidth]{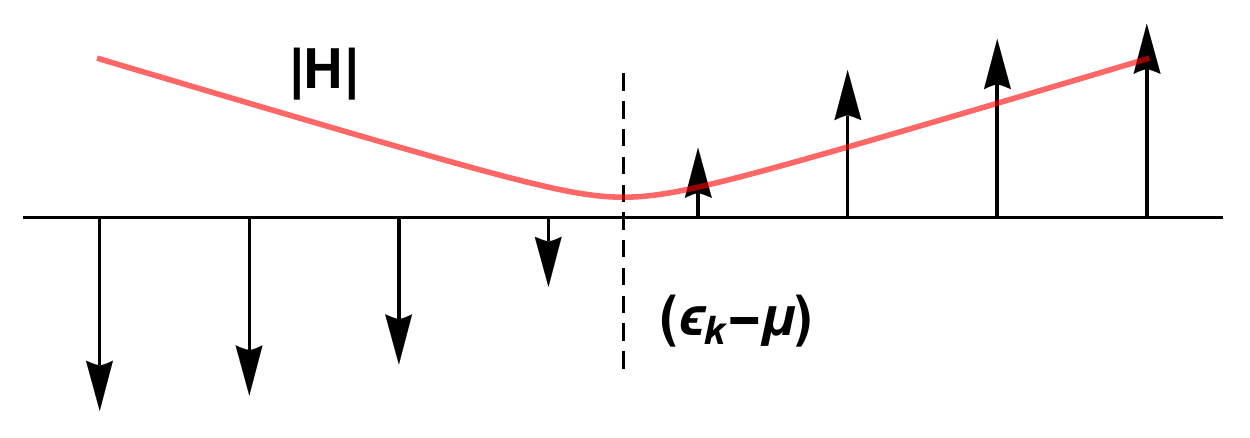}
\caption{The arrangement of spin vectors (arrows) $\mathbf{J}_\mathbf{p}$ in Anderson's reformulation of BCS theory from Ref.~\onlinecite{andersonPR58}.  The upper figure is the gapless case $\Delta=0$ with single particle energy shown by the positively sloped line.  The lower figure shows the gapped case ($\Delta\neq0$).\label{andersonFig}}
\end{figure}

The physics of this model can be understood in terms of a spin chain (see Fig.~\ref{andersonFig}). Below (above) the Fermi surface, the vectors point down (up) if there is no pairing potential.  When the pairing potential is turned on, a domain wall appears for the spin-$\frac12$ chain.  Spin up states (occupied) correspond to one type of neutrino ($\nu_1$) while spin down states correspond to the other ($\nu_2$).

Note that the magnetic field $\mathbf{H}$ is not physically applied to the electrons in the condensed matter system, it is a mathematical construction.  Adding a magnetic field $B$ into the model introduces a spin-dependent $\epsilon_F$ which shifts the vertical dashed lines in Fig.~\ref{andersonFig} for each spin.  This is understood in the neutrino case as an increase in one neutrino type and a decrease in another which is consistent with the CP violation induced by the FFLO effect (see Sec.~\ref{CPviolation}).

\end{appendix}

\bibliography{neutrino}

\end{document}